\documentclass[aps,prl,superscriptaddress,preprintnumbers,
showpacs,legalpaper,twoside,twocolumn]{revtex4}

\usepackage{bm}
\usepackage{graphicx}
\usepackage{amsfonts}
\usepackage{amsmath}
\usepackage{amssymb}

\begin{document}

\title{Mott glass in site-diluted $S=1$ antiferromagnets
with single-ion anisotropy}

\author{Tommaso Roscilde}
\affiliation{Max-Planck-Institut f\"ur Quantenoptik, Hans-Kopfermann-strasse 1,
85748 Garching, Germany}
\affiliation{Department of Physics and Astronomy, University of Southern
California, Los Angeles, CA 90089-0484}
\author{Stephan Haas}
\affiliation{Department of Physics and Astronomy, University of Southern
California, Los Angeles, CA 90089-0484}


\begin{abstract}
The interplay between site dilution and 
quantum fluctuations in $S$=$1$ Heisenberg antiferromagnets
on the square lattice is investigated using quantum 
Monte Carlo simulations. Quantum fluctuations are tuned by 
a single-ion anisotropy $D$.
In the clean limit, a sufficiently large $D>D_c = 5.65(2) J$
forces each spin 
into its $m_S=0$ state, and thus destabilizes antiferromagnetic
order. 
In the presence of site dilution, quantum fluctuations are 
found to destroy N\'eel order \emph{before} the percolation 
threshold of the lattice
is reached, if $D$ exceeds a critical value
$D^* = 2.3(2) J$. 
This mechanism opens up an extended quantum-disordered
\emph{Mott glass} phase on the percolated lattice, characterized by a
gapless spectrum and vanishing uniform susceptibility.

\end{abstract}
\maketitle

Strongly interacting quantum systems on random lattices
offer the possibility of realizing genuine quantum
phases with unconventional properties \cite{Fisheretal89}. 
The inhomogeneous nature of such disordered lattices can lead 
to the existence of strongly correlated regions with 
excitations of arbitrarily low energy, and yet 
an overall absence of long range order. Therefore quantum ``glassy''
phases are possible which, at the same time, have a \emph{quantum-disordered} 
ground state and a finite low-energy density
of states, not associated with any Goldstone mode 
\cite{Fisheretal89,Orignac,ProkofevS04}.    

A large variety of physical systems potentially
lends itself to realize the theoretical scenario of
quantum glassy phases and quantum phase transitions
from such phases into more conventional 
long-range ordered ones. Relevant
examples are $^4$He films on 
disordered substrates \cite{Csathyetal03}, 
type-II superconductors with columnar
defects \cite{SC} and trapped 
cold atoms in disordered optical potentials
\cite{coldatoms}. Unfortunately, only
in few cases has possible evidence
for quantum glassy phases
been observed, and the agreement with theoretical
predictions is still unsatisfactory. 

In this paper we numerically demonstrate
the emergence of a highly non-trivial interplay
between lattice randomness and quantum fluctuations
in low-dimensional antiferromagnets. 
Disorder can typically be introduced in antiferromagnets
by dilution of the magnetic lattice with non-magnetic
ions, and an extremely high level of control
on the concentration of dopants can be reached \cite{Vajketal02}.
Recent experiments on site diluted compounds with antiferromagnetic
interactions have shown that the destruction of magnetic 
long-range order (LRO) due to doping occurs \emph{at} the percolation 
threshold of the lattice \cite{Vajketal02, Birgeneauetal80}. 
In the case
of the $S=1/2$ square-lattice Heisenberg antiferromagnet
this is clearly understood since this particular system 
is too far from a quantum critical point to develop quantum critical 
fluctuations triggered by site dilution
\cite{VajkG02}. In this paper, we consider
a realistic magnetic model on the square lattice
in which the strength 
of quantum fluctuations can be tuned at will, driving
the system from a classically ordered state to 
a quantum disordered one. Non-linear
quantum fluctuations enhanced by lattice disorder give rise
to a novel quantum disordered phase which is 
\emph{gapless} and has a
\emph{vanishing uniform susceptibility}. An exact bosonic
mapping of the model allows us to identify this regime 
with a \emph{Mott-glass phase} \cite{ProkofevS04}
of a commensurately filled lattice gas of interacting bosons. 
The application of a magnetic field to such a phase drives it
into a disordered \emph{Bose-glass phase} \cite{Fisheretal89}, 
which maintains the gapless spectrum but acquires a finite 
susceptibility.  
 
   We investigate the $S$=$1$ square-lattice \emph{anisotropic} 
Heisenberg antiferromagnet (SLAHAF) with site dilution, 
whose Hamiltonian reads
\begin{equation}
{\cal H} = J \sum_{\langle ij \rangle} \epsilon_i \epsilon_j 
{\bm S}_{i}\cdot{\bm S}_{j} ~~+ D \sum_{i} \epsilon_i (S^z_{i})^2
- hJ \sum_{i} \epsilon_i S^z_{i}.
\label{e.Hamilton}
\end{equation}

Here $\bm S$ denotes $S$=$1$ spin operators, and 
$\langle ij \rangle$ enumerates 
pairs of nearest neighboring sites on the square lattice.
The variables $\epsilon_{i}$ take the values 0 or 1
with probability $p$ and $1-p$ respectively, where $p$
is the concentration of non-magnetic dopants. 
We make use of the Stochastic Series Expansion Quantum
Monte Carlo (SSE-QMC) method based on the operator-loop algorithm
\cite{Syljuasen03}, which allows us to faithfully
monitor the $T=0$ physics by a $\beta$-doubling approach
\cite{Sandvik02} on $L\times L$ lattices with 
$L$ up to 36 sites. The results are typically averaged
over at least 200 disorder realizations.
  
  In the clean limit $p=0$, and at zero magnetic field $h=0$,
the SU(2) symmetric version of this model for $D=0$
is known exactly to show  N\'eel LRO
\cite{NevesP86}. The presence of finite single-ion anisotropy
($D>0$) reduces the symmetry to U(1), and antiferromagnetic
ordering takes place in the $xy$ plane. In the limit $D/J\gg 1$
the ground state becomes 
$|\Psi_0\rangle = \prod_i |m_S=0\rangle_i$ to minimize the 
anisotropy term. Such a state has no antiferromagnetic
order, since all spin-spin correlation functions are
simply zero. Therefore, at a critical 
ratio $(D/J)_c$ a quantum phase transition \cite{WangW05}
occurs between the $xy$-ordered
regime and a gapped spin-liquid state with short-range 
correlations. Our QMC simulations of Eq.~(\ref{e.Hamilton})
with $h,p=0$ provide an estimate of $(D/J)_c = 5.65(2) J$,
obtained by linear scaling of the correlation length 
for the $x$($y$) spin components $\xi^{xx(yy)}\sim L$.
The quantum-critical scaling of the static structure factor 
$S^{xx(yy)}(\pi,\pi) \sim L^{\gamma/\nu-z}$ at the above
point satisfies the expected $3d$ XY universality class
with $\gamma\approx 1.32$, $\nu\approx 0.67$ and $z=1$
\cite{PelissettoV02}.
 
A better insight into the nature of the quantum phase
transition tuned by the anisotropy is 
achieved by exactly mapping the spins
onto bosons through the Holstein-Primakoff (HP) transformation
$S_i^{+} = \sqrt{1-n_i/2} ~b_i$ and $S_i^z = 1- n_i$
(together with a $\pi$ rotation of one of the two sublattices),
which gives
\begin{eqnarray}
&&{\cal H} = 
- \sum_{\langle ij \rangle} \frac{J_{ij}}{2}
\left[ \sqrt{1-\frac{n_i}{2}}~ b_i b_j^{\dagger} ~ 
\sqrt{1-\frac{n_j}{2}} 
+ {\rm h.c.} \right] \nonumber \\
&&+ \sum_{\langle ij \rangle} J_{ij} (n_i-1)(n_j-1) 
+ D \sum_i \epsilon_i (n_i - 1)^2 \nonumber \\
&& - hJ\sum_{i} \epsilon_i n_i. ~~~~~~
\label{e.Bosehamilton}
\end{eqnarray}

Here $n_i = b^{\dagger}_i b_i = 0,1,2$ is the 
dynamically constrained occupation number, and 
$J_{ij} = J \epsilon_i \epsilon_j$ prevents
bosons from hopping onto doped sites and from 
experiencing repulsion from those sites. 
It is evident that, apart from the square-root
terms in the hopping Hamiltonian, the model of
Eq.~(\ref{e.Bosehamilton}) is a 
\emph{Bose-Hubbard model} with soft-core
interactions, allowing for up to 2 particles per site. 
For $h=0$ the $Z_2$ symmetry of the spin model 
translates into a particle-hole symmetry of the 
bosonic model, which implies that the system 
is exactly \emph{half-filled}, 
$\langle n_i \rangle = 1$. Interestingly, this
still holds true in the presence of site dilution, 
which does not alter the particle-hole symmetry of
the Hamiltonian. In particular, the anisotropy
term $D$ of the spin model translates into
an \emph{on-site repulsion} for the HP bosons,  
such that the transition driven by increasing
$D/J$ can be understood as a superfluid-to-Mott-insulator
(SF-MI) transition
driven by the ratio between repulsion and hopping
\cite{Fisheretal89}. The observed $(d+1)$-XY universality
class is consistent with the commensurate filling
of the lattice \cite{Fisheretal89}. 
 
\begin{figure}[h]
\begin{center}
\includegraphics[
     width=80mm,angle=0]{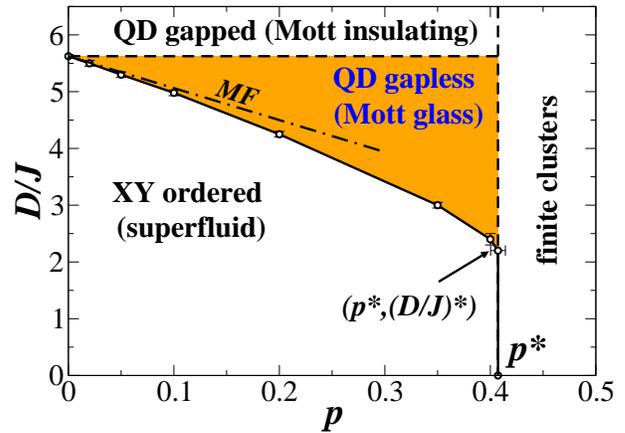} 
\caption{\emph{(color online)} Phase diagram of the site-diluted $S$=$1$ SLHAF.
The magnetic phases (and the corresponding bosonic
ones) are indicated. QD = quantum disordered.
The mean-field (MF) line corresponds to the 
low-dilution limit $(D/J)_{c,p} = (D/J)_{c,0}~(1-p)$.}
\label{f.phdiagr}
\end{center}
\end{figure}

Introducing site dilution into the Hamiltonian 
Eq.~(\ref{e.Hamilton}) hence offers the intriguing
opportunity of studying disorder effects on a
SF-MI transition, both at commensurate ($h=0$)
and at incommensurate fillings ($h\neq 0$). 
In this paper, we mainly focus on the $h=0$
regime, and we will discuss the case of $h\neq0$ in a
forthcoming publication \cite{RoscildeHnext}. 
Moreoever, the model of Eq.~(\ref{e.Hamilton})
lends itself to 
studying the behavior of antiferromagnetic 
order around the percolation threshold of
the site-diluted square lattice $p=p^*=0.407253...$
\cite{NewmanZ00}
under continuous tuning of the strength
of quantum fluctuations. In the limit $D=0$,
it has been demonstrated numerically that 
the $S$=$1$ SLHAF on a site-diluted lattice
retains LRO up to the
percolation threshold 
\cite{Katoetal00}. The system 
at finite $p$ and finite $D$ interpolates
then between the $3d$-XY transition at $p=0$
and the percolation transition at $D=0$.
 
In Fig. \ref{f.phdiagr}, we show the complete
phase diagram of the system in the $p-D$
plane at $T=0$ and $h=0$. The boundary lines
are estimated as above, using the criterion
$\xi^{xx(yy)}\sim L$. This phase diagram suggests
that the interplay between
disorder and quantum fluctuations 
leads to a significant
departure both from the SF-MI transition at $p=0$
and from the percolation driven transition at $D=0$. 
Let us start from the limit $D=0$, and 
focus on the behavior of the system 
at the percolation threshold $p=p^*$.
Here we clearly observe that, upon increasing
the ratio $D/J$, the transition line 
from magnetic LRO to disorder passes
through a multi-critical point at $(D/J)^* = 2.3(2)$,
beyond which it departs from the 
percolation threshold $p=p^*$ and bends towards
systematically smaller $p$ values 
for larger $D/J$. Starting from 
the opposite limit $p=0$, we observe
that an infinitesimal amount of disorder
leads to a shift to lower values of the critical anisotropy
$(D/J)_c$ that causes the destruction of LRO. 
Putting together these two
pieces of information, we can conclude that
for \emph{any} finite disorder concentration 
$p$ there is a non-trivial \emph{disorder-dependent}
value of the 
anisotropy $(D/J)_{c,p} < (D/J)_{c,0}$ at which 
non-linear quantum fluctuations, enhanced by 
the reduced connectivity of the diluted lattice,
are able to destroy the LRO.
 
\begin{figure}[h]
\begin{center}
\includegraphics[
     width=72mm,angle=0]{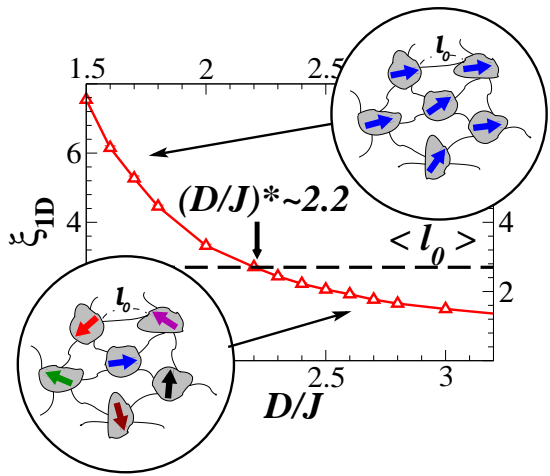} 
\caption{\emph{(color online)} Ground-state correlation length of the $S$=$1$ 
anisotropic Heisenberg antiferromagnet on the chain
as a function of the anisotropy. The illustrations
sketch the ordered/disordered phase on the percolating
cluster when the 1D correlation length is larger/smaller
than the average link length $\langle l_0 \rangle$
(see text). The arrows on the $2d$ blobs indicate
the local staggered moment 
$\sum_{i\in{\rm blob}} (-1)^i \langle S_i^{x(y)} \rangle$
or, alternatively, the local phase of the bosons.}
\label{f.xi1d}
\end{center}
\end{figure}  
 
The interplay between
disorder and quantum fluctuations can
be quantitatively understood in the relevant
limit of $p=p^*$. At this point, 
the percolating nature of the largest
cluster in the system crucially depends on 
1$d$ \emph{links} connecting
quasi-2$d$ islands with higher local 
connectivity (\emph{blobs}) \cite{Coniglio82}. 
For $D/J$, well 
below $(D/J)_{c,0}$ the quasi-2$d$ blobs
have a well-defined staggered moment 
(see Fig.~\ref{f.xi1d})    
which fluctuates with a characteristic frequency
given by the finite-size gap of the blob
if the blobs are disconnected.
When the blobs are connected by links
to form the percolating cluster, their 
local staggered moments are either long-range
correlated or not, depending on the
decay of correlations along the 1$d$ 
links \cite{BrayAlietal06}.
If the characteristic length $l_0$ of the
1$d$ links is shorter than the correlation 
length $\xi_{1d}$ for the $S=1$ anisotropic
Heisenberg model of Eq.~(\ref{e.Hamilton}) on a chain, 
the links are able to establish
2$d$ LRO in the system. Otherwise
the blobs are uncorrelated, 
and the system enters a quantum-disordered
phase. 
An earlier detailed study \cite{Coniglio82}
of the geometry of the percolating cluster  reports 
an average link length $\langle l_0 \rangle \approx 2.7$
for site percolation on the square lattice.
We have studied the correlation properties of
the anisotropic Hamiltonian Eq.~(\ref{e.Hamilton})
on a chain with SSE-QMC, systematically estimating
the one-dimensional equal-time correlation length 
$\xi_{1d}$ as a function of the anisotropy
$D/J$. According to the above argument, the 
loss of LRO on the percolating
cluster should occur for $(D/J)^*$ such that 
$\xi_{1d}[(D/J)^*] \approx \langle l_0 \rangle $.
This criterion leads to the estimate $(D/J)^*\approx 2.2$
(Fig.~\ref{f.xi1d}),
which is in excellent agreement with the location
of the multicritical point estimated by QMC, as shown in Fig. 1. 
This means that we quantitatively understand 
the deviation of the magnetic transition 
from percolation in terms of a critical enhancement 
of local quantum fluctuations on the weak links
of the percolating cluster. 
  
When $p\lesssim p^*$,
the links connecting quasi-2$d$ islands
acquire a higher connectivity, evolving from single
chains to decorated chains or ladders. The 
critical value $(D/J)_{c,p}$ for $p\lesssim p^*$ 
corresponds to the one at which the correlation
length on the \emph{quasi}-1$d$ links becomes
comparable with their average linear size, analogous 
to what happens at $p=p^*$. For weak dilution 
$p\ll p^*$, on the other hand, the disordered lattice
can be approximately represented as a homogeneous
lattice with reduced effective coordination 
$z_{\rm eff} = z(1-p)$, corresponding to the 
Hamiltonian of Eq.~(\ref{e.Hamilton}) in which the
random variables are substituted by their average
$\langle \epsilon_i \rangle = 1-p$. In this case
the critical value $(D/J)_{c,p}$ should be linearly shifted
by the disorder with respect to the $p=0$ case,
$(D/J)_{c,p} \approx (1-p)~(D/J)_{c,0}$, 
which is in very good agreement with our data
for $p\lesssim 0.05$ (see Fig.~\ref{f.phdiagr}).

\begin{figure}[h]
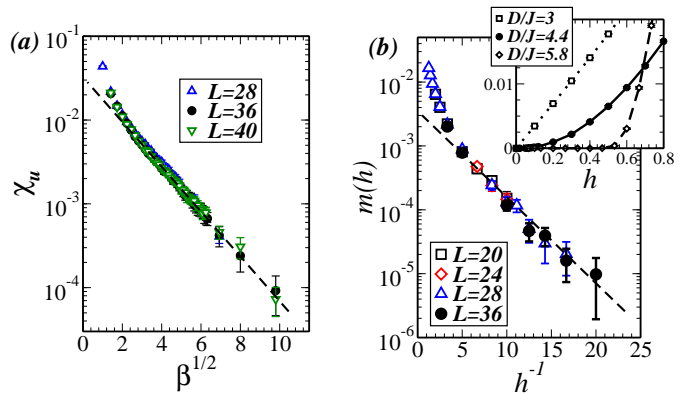

\begin{center}
\mbox{
\includegraphics[
     width=40mm,angle=0]{chi0-2Dspin1D4.4p0.2.eps} ~~~
\includegraphics[
     width=43mm,angle=0]{mhinset-2Dspin1D4.4pp0.2.eps}}      
\caption{\emph{(color online)} Magnetic response in the Mott glass phase
($p=0.2$, $D/J = 4.4$): \emph{(a)} uniform susceptibility,
\emph{(b)} uniform magnetization upon 
application of a weak field. In the inset: different 
magnetization curves starting from the XY ordered phase ($D/J = 3$),
Mott glass phase ($D/J = 4.4$), and QD gapped phase ($D/J=5.8$).
For $D/J = 3$ we have plotted $m/2$, and, for $D/J=5.8$,
$h$ is rescaled by a factor of 3. Here $L=28$ and $p=0.2$.}
\label{f.MG}
\end{center}
\end{figure}  

   The deviation of the magnetic transition from the 
percolation transition opens a novel quantum-disordered
phase that exists on percolated lattices with connectivities
arbitrarily close to the clean system as 
$D/J \to (D/J)_{c,0}$. This phase has remarkable
unconventional properties, as evidenced by 
its response to a uniform magnetic field. 
In Fig. \ref{f.MG}(a) we show the low-temperature
uniform susceptibility $\chi_u(T\to 0)$ for a 
representative point $p=0.2$, $D/J=4.4$. It is
observed that  $\chi_u(T\to 0)$ tends
to 0,
as in a phase with a singlet ground state,
so that, in the bosonic language, this phase
is \emph{incompressible} at $T=0$.
Yet the specific temperature dependence of 
$\chi_u$ appears to follow an unconventional
exponential law $\chi_u(T) \sim \exp[-\sqrt{\Delta/T}]$
for more than two decades of the QMC data.
This is clearly different from the conventional behavior
($\sim \exp[-\Delta/T]$) expected in the presence
of a finite singlet-to-triplet gap $\Delta$. To probe
this gap directly, we study the response of 
 the system to small but finite uniform fields.
Fig. \ref{f.MG}(b) shows the low-field 
magnetization curve of the system in the 
quantum disordered phase. This is strong
numerical evidence that the magnetization
is finite for arbitrarily small fields, 
although it grows slowly with $h$, 
following the unconventional exponential 
behavior $m(h) \sim \exp[-A/h]$ over
about two decades of the QMC data.
The magnetization clearly shows that the singlet-to-triplet 
gap \emph{vanishes} even at $h=0$, and this observation
is still compatible with a vanishing uniform 
susceptibility for $h=0$ because
$\chi_u = dm/dh \sim h^{-2} \exp[-A/h] \to 0$
for $h\to 0$, but it is non-zero for any 
finite $h$. The magnetization
curve in the novel disordered phase contrasts
with the linear behavior of $m(h)$ when starting from 
the XY ordered phase and with the gapped behavior 
when starting from the Mott insulating phase
(inset of Fig.~\ref{f.MG}).
The exponential $T$-dependence
of $\chi_u$ and $h$-dependence of $m$ can be
fully captured by a simple model in
which the response to a weak field is given
by independent rare clean regions of the 
percolating cluster, whose exponentially
rare statistics reflects itself in the 
exponentially small response functions
\cite{RoscildeHnext, Roscilde06}. 
The gapless nature of the spectrum is
a property of the \emph{entire} novel disordered
phase for $(D/J)_{c,p} \leq  D/J \leq (D/J)_{c,0}$.
In fact, for any $p$
one can always find an arbitrarily large (albeit
rare) clean region which locally approximates the 
behavior of the clean square lattice and which has
consequently arbitrarily small local excitations
as long as $D/J \leq (D/J)_{c,0}$.
 
   To summarize, we have seen evidence of 
an \emph{incompressible}, yet \emph{gapless}
phase of the system. The amorphous nature 
of this disordered phase and its rich low-energy 
dynamics make it akin to 
glassy phases, and at the same time its
incompressibility, along with the commensurate
filling of the lattice, bears a strong resemblance 
to a Mott insulating phase. Hence it can be called a
\emph{Mott glass}, a name that has been used in 
the recent literature for dirty-boson models  
at commensurate filling \cite{ProkofevS04}
in analogy with a phase of disordered fermions
having the same thermodynamic/spectral signatures
\cite{Orignac}. Moreover, for any finite $h$
the ground-state phase acquires a finite
susceptibility, which means that the Mott-glass
phase is driven into a \emph{Bose-glass}
\cite{Fisheretal89} which
is \emph{disordered}, \emph{gapless} and 
\emph{compressible} \cite{RoscildeHnext}.  
 
The occurrence of a bosonic 
Mott glass phase in the in site-diluted $S=1$ antiferromagnet
with single-ion anisotropy 
is well understood within the bosonic mapping
of Eq.~(\ref{e.Bosehamilton}) \cite{ProkofevS04}. In particular
we wish to stress that the Hamiltonian of
Eq.~(\ref{e.Hamilton}), both in its clean and
disordered form, is a reliable description of
the magnetic degrees of freedom in strongly
anisotropic $S$=$1$ antiferromagnetic insulators
with and without doping. 
In fact, the gapped quantum-disordered state induced by the 
large anisotropy term is realized in a variety of Ni-based
compounds, such as Ni(C$_5$H$_5$NO)$_6$(NO$_3$)$_2$ \cite{Carlinetal79},
[Ni(C$_5$H$_5$NO)$_6$](ClO$_4$)$_2$ \cite{Diederixetal77}, and the more
recently investigated NiCl$_2$$\cdot$4SC(NH$_2$)$_2$ 
\cite{Paduan-Filho04}. These systems have in general a 
three-dimensional magnetic lattice, but the results we have
shown for the 2$d$ case can be generalized straightforwardly 
to 3$d$. Moreover one can imagine tuning the $D/J$
ratio experimentally by applying hydrostatic pressure to the 
crystals. 
 
The $Z_2$ symmetry of the magnetic Hamiltonian in zero 
field for all these systems
translates into particle-hole symmetry
and commensurate (half) filling of the
soft-core bosons on \emph{any lattice site}, 
regardless the geometry
of the lattice itself, namely also on 
random percolating clusters. 
This ingredient, which is crucial 
for the appearance of the Mott glass,
appears hard to be achieved in systems of
real bosons, as \emph{e.g.} in bosons in optical 
lattices, where the introduction of 
disorder will invariably change the 
particle occupation near lattice defects.

We acknowledge fruitful discussions with N. Bray-Ali,
J. Moore, B. Normand, N. Prokof'ev, P. Sengupta,
and I. Zaliznyak. 
This work is supported by the DOE
through grant No. DE-FG02-05ER46240, and by the European
Union through the SCALA integrated project. 
Computational facilities
have been generously provided by the HPCC-USC Center.

\end{document}